\documentclass{elsart}

\usepackage{natbib}

\usepackage{graphicx}

\usepackage{amssymb}

\begin{document}

\begin{frontmatter}



\title{Astrometry with the Keck-Interferometer: the ASTRA project and its science}

\author[label1,label2]{Jorg-Uwe Pott}
\author[label1]{Julien Woillez}
\author[label3]{Rachel L. Akeson}
\author[label1]{Ben Berkey}
\author[label4]{Mark M. Colavita}
\author[label1]{Andrew Cooper}
\author[label5]{Josh A. Eisner}
\author[label2]{Andrea M. Ghez}
\author[label6]{James R. Graham}
\author[label7]{Lynne Hillenbrand}
\author[label1]{Michael Hrynewych}
\author[label1]{Drew Medeiros}
\author[label3]{Rafael Millan-Gabet}
\author[label8]{John Monnier}
\author[label1]{Douglas Morrison}
\author[label1]{Tatyana Panteleeva}
\author[label6]{Eliot Quataert}
\author[label1]{Bill Randolph}
\author[label1]{Brett Smith}
\author[label1]{Kellee Summers}
\author[label1]{Kevin Tsubota}
\author[label1]{Colette Tyau}
\author[label6]{Nevin Weinberg}
\author[label1]{Ed Wetherell}
\author[label1]{Peter L. Wizinowich}

\address[label1]{W.\,M. Keck Observatory, Kamuela, Hi 96743, USA}
\address[label2]{Dept. of Astronomy, University of California Los Angeles, CA 90095, USA}
\address[label3]{NASA Exoplanet Science Institute, Caltech, Pasadena, CA 91125, USA}
\address[label4]{Jet Propulsion Laboratory, California Institute of Technology, Pasadena, CA 91109, USA}
\address[label5]{Steward Observatory, University of Arizona, Tucson, AZ 85721, USA}
\address[label6]{Astronomy Department, University of California Berkeley, CA 94720, USA}
\address[label7]{California Institute of Technology, Pasadena, CA 91125, USA}
\address[label8]{University of Michigan, Ann Arbor, MI 48109, USA}

\begin{abstract}
The sensitivity and astrometry upgrade ASTRA of the Keck Interferometer is introduced. After a brief overview of the underlying interferometric principles, the technology and concepts of the upgrade are presented. The interferometric dual-field technology of ASTRA will provide the KI with the means to observe two objects simultaneously, and measure the distance between them with a precision eventually better than 100~$\mu$as. This astrometric functionality of ASTRA will add a unique observing tool to fields of astrophysical research as diverse as exo-planetary kinematics, binary astrometry, and the investigation of stars accelerated by the massive black hole in the center of the Milky Way as discussed in this contribution.

\end{abstract}

\begin{keyword}

 Planets \sep Binaries \sep Galactic center \sep Dual-field Interferometry \sep Astrometry 
\end{keyword}

\end{frontmatter}

\section{Introduction}
\label{sec:1}

\begin{figure}[htp]
\centering
\includegraphics[width=0.52\textwidth]{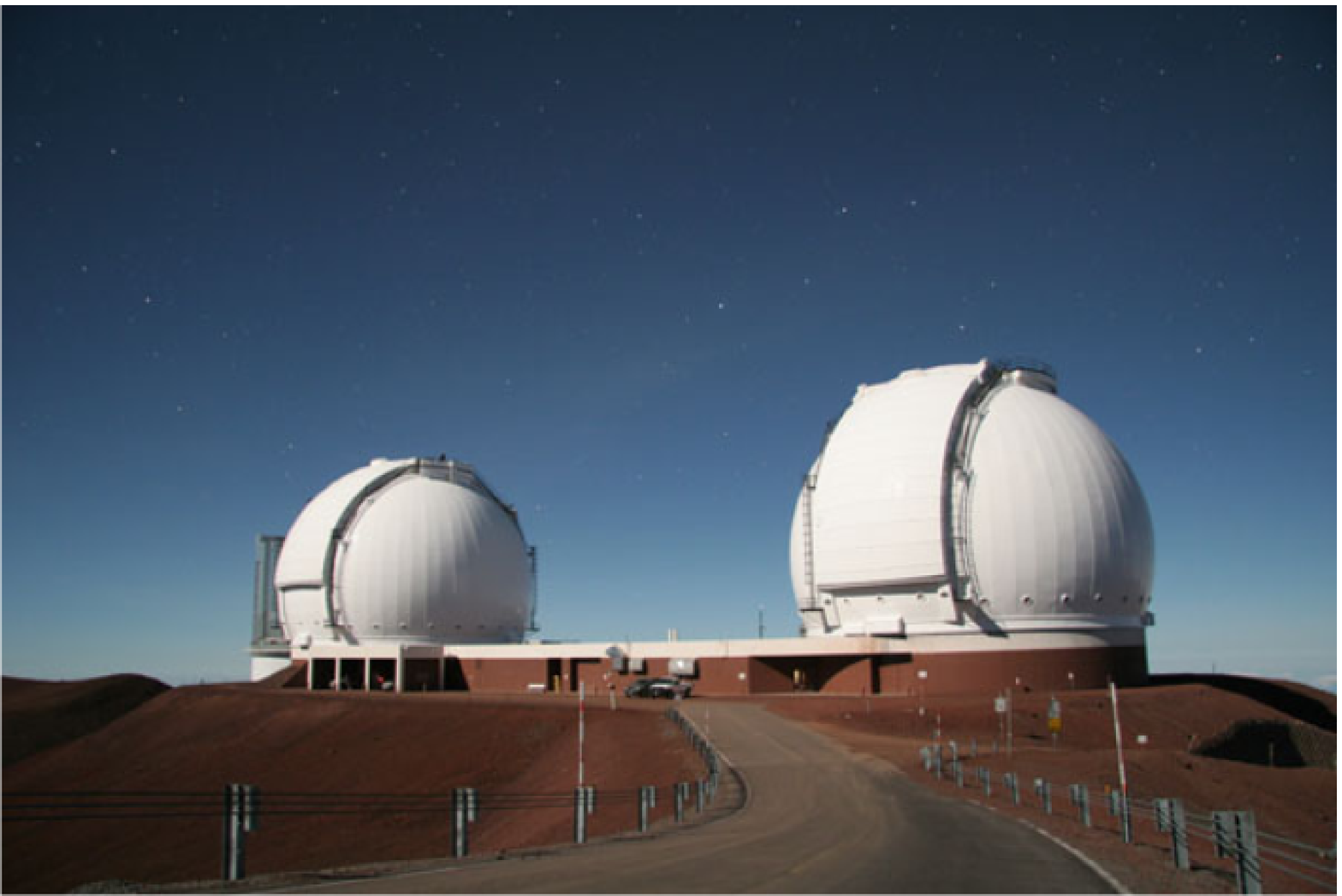}
\includegraphics[width=0.46\textwidth]{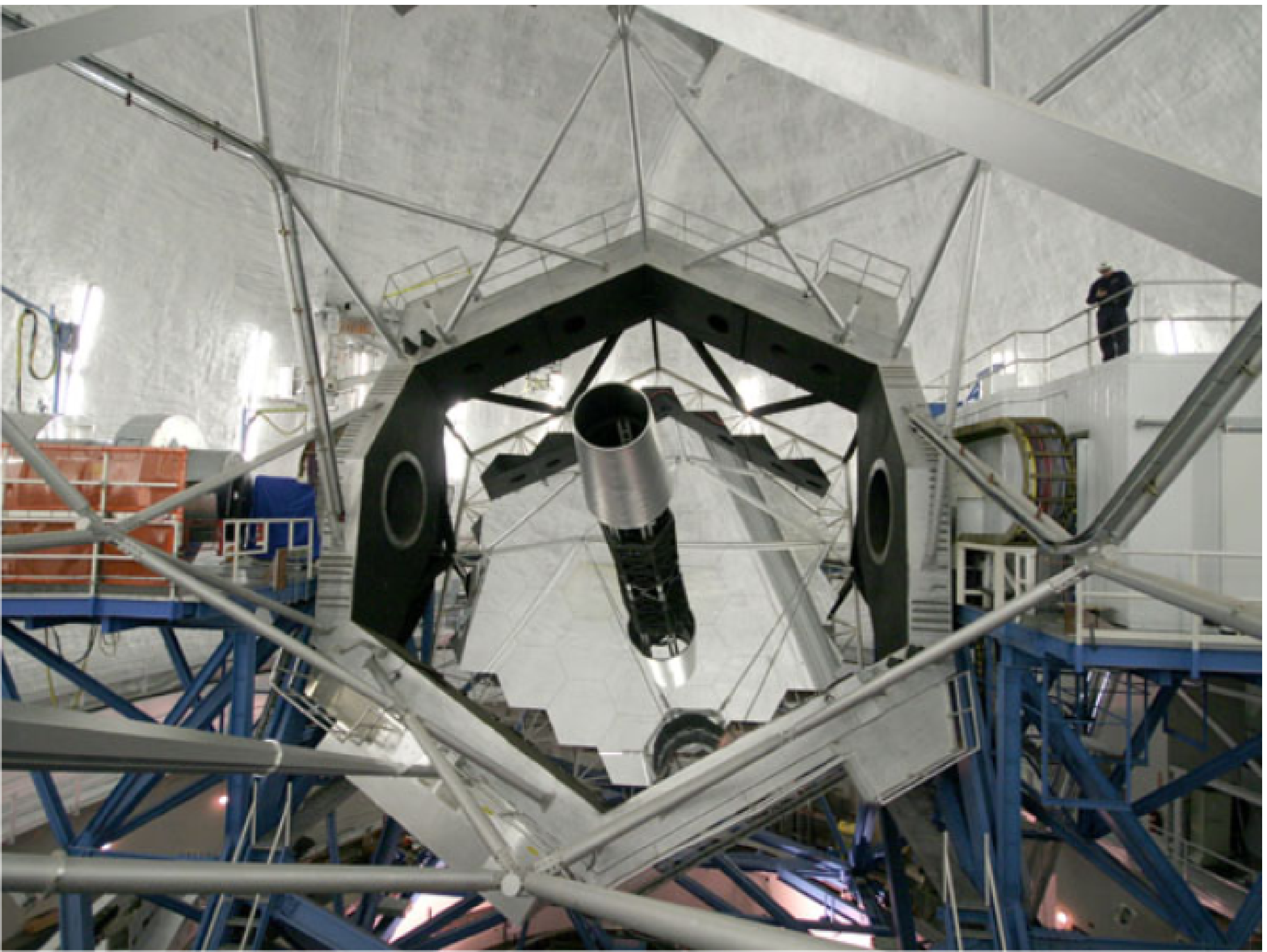}
\caption{\label{fig:11}{\it Left}: The domes of the two telescopes of the W.\,M.~Keck Observatory atop Mauna Kea, Hawaii. The telescopes are separated by 85~m and make up the Keck Interferometer. {\it Right:} One of the segmented 10~m primary mirrors.}
\end{figure}

The Keck Interferometer (KI, Fig.~\ref{fig:11}) combines the two 10m Keck telescopes with a baseline separation of 85m to simulate a large telescope diameter in terms of angular resolution. The resulting resolution of 5~mas at 2.2~$\mu$m is slightly better than the diffraction limit of the $\sim\,30$~m diameter next generation of ground-based telescopes currently under development. 
Long baseline interferometers prevail in areas of research where  the highest angular resolution is required together with only limited demands on sensitivity and imaging information, e.g. such as resolving circumstellar shell and disk emission to understand the formation of gas and dust therein.
Co-phased arrays provide already today a glimpse onto discoveries of the next generation of large ground-based telescopes.
In particular the interferometric narrow-angle astrometry, the central topic of this workshop, is a field with  long-term prospects for interferometric research.
The opto-mechanical complexity of the imaging process of 30~m telescopes \citep[see][for current plans]{2008SPIE.701219G} will make it difficult to match and outperform the high precision of {\it interferometric} astrometry with {\it imaging} astrometry.

The KI is one of the two large aperture optical long baseline interferometric (OLBI) facilities in the world. 
Funded by NASA, the KI is developed and operated by JPL\footnote{{\bf J}et {\bf P}ropulsion {\bf L}aboratory; {\tt http://planetquest.jpl.nasa.gov/Keck/keck\_index.cfm}}, NExScI\footnote{{\bf N}ASA {\bf Ex}oplanet {\bf Sc}ience {\bf I}nstitute, the former Michelson Science Center (MSC) is an integral part of NExSci; {\tt http://nexsci.caltech.edu}} and the W.~M. Keck Observatory (WMKO\footnote{http://keckobservatory.org}).  
The KI has been utilized for studying a range of astrophysics, including young stellar object disks and the first infrared interferometry observations of an AGN. 
Recent developments include the addition of nulling interferometry and improved sensitivity \citep{2008SPIE.70130AC}. For more information about using KI, its current limiting magnitudes, and target requirements, see the KI performance information at the NExScI/MSC support page\footnote{http://msc.caltech.edu/software/KISupport}.

A new major development effort is underway to broaden the astrophysical applications of this unique instrument: the ASTRA upgrade.
ASTRA stands for the ASTrometric and phase-Referenced Astronomy upgrade of the KI project. ASTRA is funded by the National Science Foundation (NSF) Major Research Instrumentation (MRI) program and will be implemented in three steps over the next 2 years. Besides the NSF engagement, a number of science institutes contribute to the ASTRA collaboration to advance and profit from large-aperture OLBI (UC Berkeley, UCLA, Caltech, NExScI, JPL,  University of Arizona).

After a conceptual introduction in Sect.~\ref{sec:2} to the underlying physical theories,
we present in Sect.~\ref{sec:3} the three modes of ASTRA, each of which is developed to overcome the current sensitivity limitations by continuous correction for phase distortions induced by the turbulent atmosphere: (i) self-referenced spectroscopy stabilizes fringes on-axis and enable higher spectral resolution up to a few thousands (ii) dual-field visibility measurements stand for integration beyond the atmospheric coherence time to reach K=15mag on science targets while locking the fringe tracker on an offset guide star (iii) the narrow-angle astrometry mode eventually will measure distances between a pair of stars within the iso-pistonic patch to a precision of better than 100~$\mu$as. Sect.~\ref{sec:4} is dedicated to describing the individual sub-systems of the ASTRA-upgrade. The science cases, which  drive the instrumentation work, are presented in Sect.~\ref{sec:5} with an emphasis on the scientific possibilities of interferometric astrometry.

\section{Theory of dual-field interferometry}
\label{sec:2}
Interferometry surpasses the fundamental resolution limits of imaging astronomy with single telescopes by probing higher spatial frequencies \citep[see other lectures in these series, e.g.][]{2007NewAR..51..565H}. While current ground-based telescope diameters, defining the angular resolution of the telescope, are limited to 8-10~m, depending on the exact design of the telescope, the interferometric angular resolution scales with the baseline, the distance between the co-phased telescopes. The current technical baseline length limitation for ground-based interferometric arrays is 1-2 orders of magnitude larger than the single telescopes diameter limit, explaining the superior angular resolution of an interferometer. 
Going to very long beam trains and delay lines does not only bear technical difficulties, but is also challenging from the astronomical point of view: An interferometer is most sensitive to sources which are of the size of the nominal resolution of the interferometer ($\lambda_{\rm obs}\,B_{\rm proj}^{-1}$, $^{\ref{foot:1}}$) or smaller. That means the larger the baseline is, the smaller the source needs to be to produce a strong instrument response (remember the similarity theorem of the Fourier transform). But smaller sources are usually farther away and thus apparently fainter. 
In other words, sources which are so small that they would require kilometer-long baselines to be resolved are usually so faint that very large individual apertures are needed to deliver enough photons
\footnote{This is a difference to radio interferometry. A lot of very bright radio sources are very powerful non-thermal synchrotron sources in quasars which have source sizes by orders of magnitude smaller than any reasonable thermal sources of similar flux. The infrared sky however is dominated by thermal sources: stars and the surrounding, heated matter.
}.

Dual-field interferometry and phase-referencing was invented to cope with the sensitivity limiting effects of the turbulent atmosphere, which so far has limited the application of the resolution advantage of long baseline interferometers in optical- and near-infrared astronomy. 
Thus dual-field interferometry does not reach beyond the fundamental resolution limits of interferometry but it helps reaching these limits on fainter objects in the reality of a ground-based telescope array. 
To understand this we focus on the visibility phase for a moment.

\subsection{Phases in interferometry}
\begin{figure}[t!p]
\centering
\includegraphics[width=0.8\textwidth]{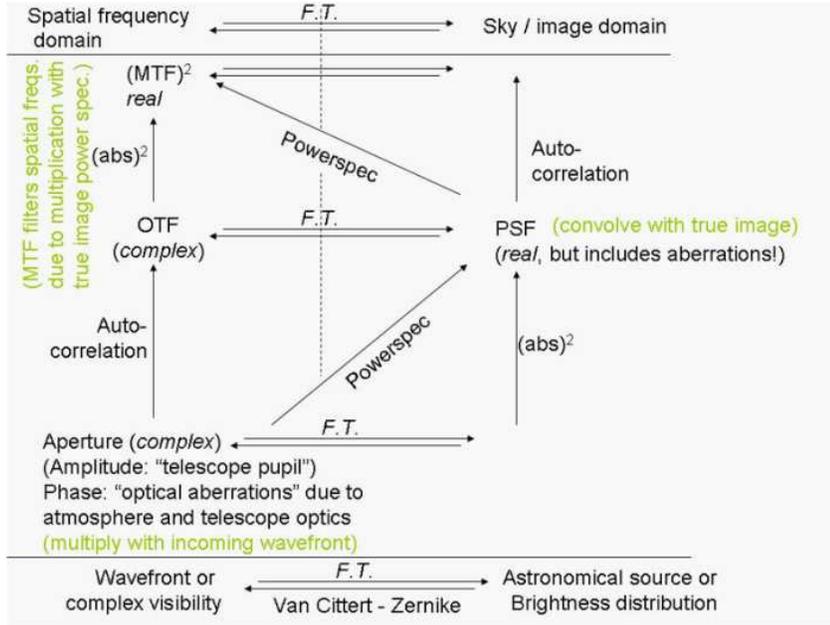}
\caption{\label{fig:20}Sketch describing how the Fourier transform connects the individual elements of an astronomical imaging process: aperture, optical transfer function (OTF), point spread function (PSF), modulation transfer function (MTF). It becomes obvious that the telescope aperture is a filter for certain spatial frequencies. Here {\it telescope} can similarly mean the circular aperture of a single telescope or the spotted aperture of a co-phased array.}
\end{figure}

\begin{figure}[tp]
\centering
\includegraphics[]{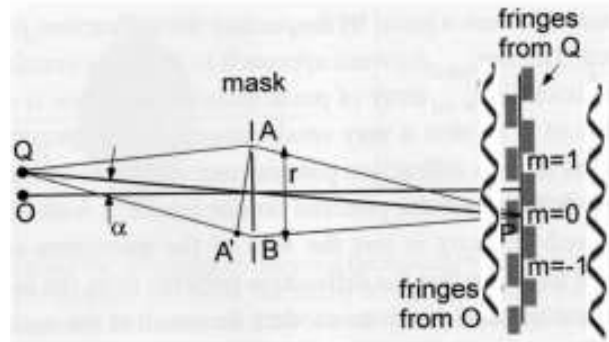}
\caption{\label{fig:21}Two-element interferometer seen as a double-slit experiment. Here the connection between {\it phase} of the fringe pattern and fringe location becomes visible. The target is simplified to two point sources $Q$ and $O$, separated by the angle $\alpha$. The optical path difference (OPD) equals AP-BP. And the differential OPD is geometrically connected to the angle $\alpha$ and to the difference in fringe positions, each given by the location of the intensity maximum, denoted with m=0 for the fringe from Q. This is the basis of interferometric astrometry. The slit distance $r$ stands for the baseline. The angular distance $\alpha$ can be derived from the differential fringe position and the baseline length \citep[compare also to Eq.~\ref{eq:1} in the text, the sketch is derived from][]{2006iosi.book.....L}.}
\end{figure}

An interferometric measurement aims at measuring the amplitude and phase of the target at a certain spatial frequency
\footnote{\label{foot:1}The spatial frequency probed is given by the ratio $B_{\rm proj}\,\lambda^{-1}_{\rm obs}$ and has the dimension of inverse radians. $B_{\rm proj}$ is the projected baseline length at the moment of observation, and $\lambda_{\rm obs}$ denotes the effective observing wavelength. The reciprocal of the spatial frequency, $\lambda_{\rm obs}\,B_{\rm proj}^{-1}$ is the nominal resolution of the interferometer.}.
Measuring the amplitude or power at this frequency is typically easier than measuring the phase.
While the amplitude is a measure for the compactness of the source, the phase of the interferometric observable, the visibility, contains crucial information about the brightness {\it distribution} of the target, that is the details of the source structure. 
The phase of the spatial frequency is given by the shape of the target and is independent of time until the shape changes due to an astrophysical change in the source, e.g. the creation of a new dust shell. 
This connection between interferometric phase and imaging information has been introduced by other authors in the series \citep[e.g.][]{2007NewAR..51..604M} and is most comprehensively described by the van~Cittert-Zernike theorem. It states that the brightness distribution of a target can be analyzed by a two-dimensional Fourier-transformation, whereas a simple one-baseline interferometer, a co-phased two-telescope array, is {\it the} tool to measure one of the Fourier components at a time. 
This is nothing particularly surprising, since the Fourier transform plays a crucial role in {\it every} type of astronomical imaging, as summarized in Fig.~\ref{fig:20}.

The connection of visibility phase and imaging information can be understood immediately by acknowledging the role of the Fourier transform. Translating the position in the image space (that is "moving the star on the sky") is not changing the amplitude in visibility space (because the target still "looks the same"), but a linear phase slope is added to the complex visibility function. 
The meaning of the visibility phase as a way to locate the source in the sky with respect to a deliberately chosen phase center (typically a nearby bright star) is shown in Fig.~\ref{fig:21} by the analogy between a two-element interferometer and imaging through a double-slit. 

The interferometric phase or visibility phase basically encodes the optical path differences between two incoming rays of light. This leads immediately to the understanding of how the turbulent atmosphere hampers interferometric phase measurements: Different optical path lengths or different indices of refraction along different ways through the atmosphere are created by continuous temperature, density and compositional variations of the atmosphere. While temperature and density changes (e.g. due to winds) dominate the variation of the atmospheric index of refraction in the visible, the humidity becomes more and more important toward the longer wavelengths in the infrared. In particular the {\it changes} of humidity can be rather large in a short time due to the non-homogeneous distribution of water in the atmosphere (think of clouds).

\begin{figure}[htp]
\centering
\includegraphics{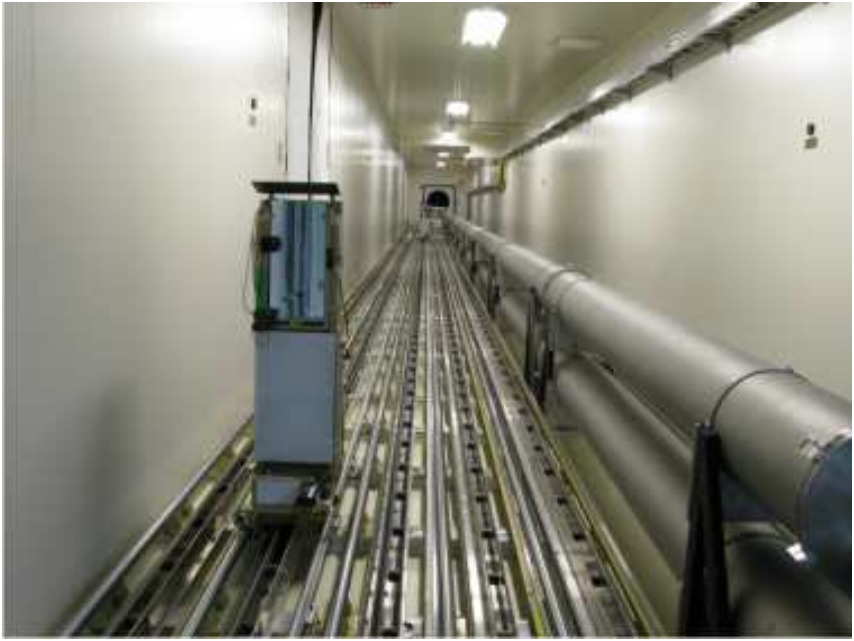}
\caption{\label{fig:1}The quasi-static long delay lines of the Keck Interferometer can correct for $\pm\,70\,{\rm m}$ of geometric delay.}
\end{figure}

The visibility phase of a target is only defined with respect to a phase center. This can be given by the fringe position
\footnote{\label{foot:9}Note that {\it fringe position} usually refers to the OPD of the starlight, which has to be equalized by the delay lines to enable interference (Fig.~\ref{fig:21}
).
In the case of the KI, the delay lines consist of movable mirrors of which the actual position is known and continuously measured. 
Fig.~\ref{fig:1} shows the long delay lines of the KI, which are a part of the delay line system. 
It is convenient to think of the fringe position as the physical position of the delay line reflectors, at which the center of the interferometric fringe pattern is found  \citep[see Fig.~4 in][]{2007NewAR..51..565H}. However similarly to the phase, this is only a differential measure, and the zeropoint is arbitrary.
}
 of another star, or by the fringe position of the same object but observed at a different baseline. The difficulty is to ensure that the measured phase, as a differential information, only traces the true geometrical delay, and is not altered by the turbulent atmosphere or instrument-induced phase changes.

Here we need to discuss two properties of the turbulent atmosphere which dominate the difficulty of a phase measurement, and therefore an interferometric measurement at all: piston stability (Sect.~\ref{sec:22}) and  an-isopistonism (Sect.~\ref{sec:23}.

\subsection{Piston stability} 
\label{sec:22}

The atmosphere adds a random phase to each beam due to continuously varying index of refraction. This phase, averaged over the telescope aperture, is called {\it piston}. 
Being constant over the aperture, the piston does not affect single telescope imaging (Fig.~\ref{fig:20}), but a differential piston between two telescopes changes the effective OPD and biases a phase (= astrometric) measurement.

Although piston is only seen in interferometric data, it goes back to the same atmospheric turbulence responsible for the seeing. Therefore the magnitude and velocity of the differential piston variation is roughly correlated with the magnitude of seeing and the atmospheric coherence time. 
Since the fringe position is continuously varied by the differential piston, the integration time needs to be shorter than the timescale of the piston variation to avoid a blurring of the fringe pattern which will decrease the measured fringe contrast (Eq.~\ref{eq:2}). 
Too large a phase variation during an interferometric measurement, which takes a finite amount of time and therefore averages the visibility amplitude, reduces the fringe SNR in the following way
\begin{eqnarray}
\label{eq:2}
V_{\rm avg}^2 &=& V_0\,\exp{(-\sigma _{\rm rad}^2)}
\end{eqnarray}
where $V_0$ and $V_{\rm avg}$ are the original and time-averaged visibility amplitudes, and $\sigma _{\rm rad}^2$ is the phase variance over the detector integration time.

That means, similar to speckle imaging, the signal-to-noise ratio (SNR) of the observable, the {\it correlated} flux, {\it decreases} with increasing integration time when the integration time is longer than timescale of piston stability (tens to hundreds of milliseconds in the infrared, depending on the exact observing wavelength and current atmospheric conditions). But the only way to estimate the current optimum position of the delay lines is to analyze the actual fringe. If the SNR of that fringe is not high enough to do so, the derived delay line position is faulty and the fringe is lost. 
Thus the changing atmospheric piston is limiting the integration time and the sensitivity of the interferometer. This is one of the two fundamental reasons why until today co-phased arrays are relatively insensitive and bound to study bright objects.
The other reason is the low optical throughput of an interferometer, which originates in the need for 2-3 times as many reflecting / transmitting optics as needed for a single telescope imaging camera and in intensity losses due to the large distance between the primary mirror and the detector.

The only way to overcome this sensitivity problem is a dual-field interferometer which observes at the same time two nearby stars through similar atmospheric piston, and with one of them being bright enough to measure the fringe and preset the delay lines correctly. 
This fringe stabilization enabling longer coherent integration times and higher SNR on the second star is called {\it fringe-tracking based on phase-referencing}. It is the basis of the {\sc Astra} project.

\subsection{An-isopistonism}
\label{sec:23}
Similar to the effect of an-isoplanatism, which limits the correction of an adaptive optics (AO) experiment over the field-of-view, the differential piston correction of phase-referencing (PR) degrades with the separation distance between the two fields observed. If the two stars are too far away, the turbulence profiles along the individual lines of sight to each star are not correlated to each other anymore, and one cannot correct for the piston toward the fainter star $B$ by measuring the piston toward the brighter star $A$. Star $A$ is also called the {\it phase-reference}. 
The differential piston between the two stars should not increase to 1~rad or beyond to keep the SNR losses smaller than 20~\% (Eq.~\ref{eq:2}). This limits the isopistonic angle even at the best sites (Mauna Kea, Cerro Paranal) to less than 30~arcsec at 2.2~$\mu$m \citep{2000AA...353L..29E}. 

Knowing the synonymous meaning of phase and distance between the two (unresolved) stars, relates the atmospheric residual piston noise of a differential measurement to the astrometric precision.
\citet{1992A&A...262..353S} showed that in the narrow angle regime, where the differential piston noise is small, longer baselines $B$ and smaller star-star separations $\theta$ decrease the impact of atmospheric piston on an interferometric astrometry measurement (astrometric error $\propto\,\theta\,B^{-2/3}$).
Thus longer baselines are favorable for instruments like ASTRA. Also the outer scale of the atmospheric turbulence and its relation to the baseline relates to the astrometric noise. The longer the baseline is with respect to the outer scale, the smaller the astrometric error is as derived from the differential piston noise \citep[see Fig.~3\,\&\,4 in][]{1992A&A...262..353S}.

\section{Project milestones}
\label{sec:3}
The ASTRA project aims at upgrading the KI with the technology needed to alleviate the limitations of ground-based interferometry imposed by the turbulent atmosphere. Due to the fact that the project upgrades an existing instrument used for scientific observations by the community
both the hardware testing and implementation has to occur with the least possible impact on the operating system. At the same time specifications for instrumental performance and stability are different for individual goals of ASTRA. 
To cope with this situation we designed a modular approach which develops the ASTRA functionality gradually and offers step-by-step  integration into the existing system. Lessons learned from each previous step feed back into the next, and leave some flexibility for the final design and implementation. 
The latter is necessary to efficiently realize such a complex project in which every part of the infrastructure affects the final instrument performance.
In the following the three major project phases are outlined.

\subsection{Phase~1: Self-phase-referencing}

\begin{figure}[htp]
\centering
\includegraphics[]{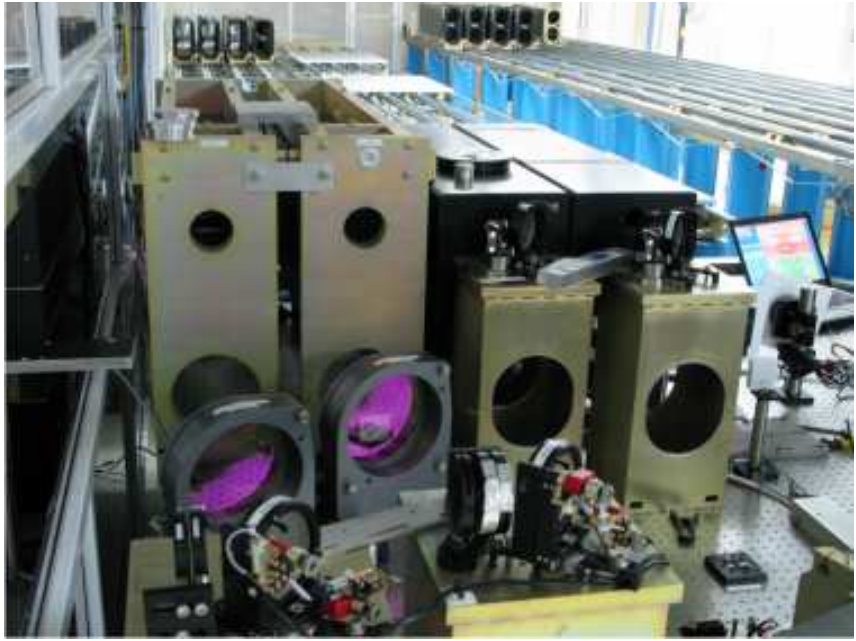}
\caption{\label{fig:10}The fast delay lines are shown in the back of the picture. They add delay to the LDL's shown in Fig.~\ref{fig:1} to complete the correction for the geometric delay. In addition they correct for atmospheric turbulence at high bandwidth (a few hundred Hz) and create the controlled fringe pattern in delay space by introducing a saw-tooth delay. In the foreground the beam splitting optics are visible which are used in the ASTRA-SPR mode to split the light before the beam combination.}
\end{figure}

A lot of physical insights derive from spectroscopy, emission lines reveal their excitation conditions, absorption lines trace the chemistry of transmitted matter etc. 
A similar line of reasoning holds for the spectroscopically resolved visibility measurement: spectro-interferometry. 
The obvious advantage of spectro-interferometry is to be able to compare the size scales of one spectral region with another, e.g. to estimate how far the line-emitting region is from the stellar photosphere. In addition, also the differential phase can be retrieved when the spectral resolution is high enough to reveal non-linear phase changes over the spectrum. Such phase shifts indicate translations of the photo-center over the respective spectral channels on the sky. For example an emission line can show a differential phase signature when emitted from an outflow. Recently \citet{2007NewAR..51..724W} studied such differential phase signals of the LBV $\eta$~Carinae with VLTI/AMBER demonstrating the astrophysical potential of such measurements. 

Due to the light dispersion and less flux per spectral channel, spectro-interferometry hits sensitivity limits even sooner than quasi-continuum measurements (see the argumentation in the previous section). 
In the $K$-band the KI reaches its sensitivity limit at $K\,\sim\,$10 (7) with a spectral resolution of $R\,\sim\,$30 (230). The ASTRA self-phase-referencing (SPR) mode breaks that limit for high dispersion spectroscopy, currently enabling $R\,\sim\,$1800 at $K\,\sim\,7$. 
Right before the beam combination the light is split and sent in parallel to two fringe cameras. 
While the first beam is dispersed only over 5 pixels to enable phase and group delay estimation for a fast fringe stabilization, the second beam passes a grism to achieve the maximum dispersion of $R\,\sim\,1800$ provided by the KI fringe cameras FATCAT \citep[see][]{2003SPIE.4838824V}. 
The first or primary fringe camera, acting as a {\it fringe tracker}, commands the delay lines to continuously take out the piston and keep the phase rms within about a radian.
Now the necessary SNR for a solid fringe detection of about 100 can be achieved at the second fringe camera in each spectral channel simply by increasing the detector integration times (DIT).
DITs as long as 1~sec and longer can be achieved in SPR as recently demonstrated in an ASTRA-commissioning run, longer by a factor of 100 or more than the usual limits imposed by the atmospheric piston noise.

\subsection{Phase~2: Dual-field operation}
\label{sec:32}
The dual-field operation is a natural extension of the SPR mode. Again two fringe cameras run in parallel. The first one tracks the fringes typically at 250~Hz for good piston and vibration correction, enabling much longer DITs at the second camera to increase the SNR. 
But in contrast to SPR, the dual-field phase-referencing mode (DFPR) focuses on increasing the limiting magnitude of the low-dispersion mode by about 5~magnitudes by pointing the second fringe camera on a faint star within the isopistonic patch around a bright star.
The new limiting magnitudes achieved by dual-field operation will open a whole new ensemble of observable targets and science cases. The key difference in the implementation between these first two phases is that for the dual-field operation the light has to be split already in the image plane at the Nasmyth foci of the telescopes. 

After this field separation (see Sect.~\ref{sec:42}) the light travels along two separated beam trains down to the beam combining laboratory, thus a doubled delay line infrastructure is needed.
The additional delay lines are already in place and in regular use for the operation of the KI-Nuller instrument, a 10~$\mu$m nulling interferometer build to detect and study the exo-zodiacal dust around nearby stars \citep{2008SPIE.70130AC}. 
The atmospheric piston, as measured on the bright star in the primary field, can be applied to both fields to stabilize the fringe motion. This correction is effective as long as the star separation is smaller than the isopistonic angle (see Sect.~\ref{sec:23}). But monitoring helper systems are needed to ensure that the non-common path after the beam separation does not suffer from vibration induced decorrelation and differential tip-tilt. Colavita (2008) discusses adverse effects in dual-feed interferometry in a dedicated contribution to these proceedings.

First on-sky tests with the dual-field operation are planned for the summer of 2009. The advantage of the dual field operation is two-fold. It will enable visibility amplitude measurements on fainter stars reaching limits well beyond $K\,\sim\,$10, a brightness range which yet has not been explored at all, and which will open the door to systematic interferometric studies of faint binary companions, YSO's and AGN. 
But dual field operation will also allow to use the one (unresolved) star as phase reference against the other. 
An unresolved star has no intrinsic visibility phase, so the measured phase derives entirely from the atmosphere and the instrument. 
This knowledge can be used to calibrate and retrieve the intrinsic phase information from the (fainter) companion in the second field. 
As mentioned before, this will help to disentangle imaging information about the object in the second field, e.g. asymmetric dust around an AGN would produce such an intrinsic phase. 

\subsection{Phase~3: Interferometric astrometry}
\label{sec:33}
Once stable dual field operation is available, an {\it astrometric} measurement between both targets can be conducted. Such an astrometric measurement is complementary to the dual-field visibility experiment described in the previous section because the interferometric astrometry focuses on measuring the differential fringe location or phase between the two fringe packets, rather than the fringe contrast in the individual packets. For the phase-referenced visibility measurement the absolute fringe location does not matter, as long as the geometrical delay is corrected for well enough to ensure that the fringe pattern stably ends up on the detector. The visibility information is encoded in the {\it contrast} of the interference signal. 
But also the differential fringe {\it location} contains astronomical information, as shown by the following equation
\begin{eqnarray}
\label{eq:1}
OPD_{\rm i}\,=\,\vec{s_{\rm i}}\cdot \vec{B},\quad{\rm and}\quad \Delta OPD = \Delta \vec{s}\cdot \vec{B}
\end{eqnarray}
where atmospheric and instrumental contributions and intrinsic source phases are neglected. $\vec{B}$ is the three-dimensional baseline vector, connecting the pivot points of the telescopes, also called the wide-angle baseline for it predicts via Eq.~\ref{eq:1} the fringe location due to the geometric delay over the entire accessible sky or over wide angles between individually observed stars.
$\vec{s_{\rm i}}$ denotes the direction toward each of the stars. 

The product of both is the geometrical optical path difference $OPD_{\rm i}$ or the fringe position in the laboratory $^{\ref{foot:9}}$. 
It relates to the amount of optical path which the delay lines have to correct for. In reality, the location and vibration of the mirrors and the current index of refraction of the atmosphere additionally affect the exact location of the fringes. But if those effects are stable and comparable for both stars they cancel out in the differential measurement, and only the true geometric difference survives as written down in the second part of Eq.~\ref{eq:1}. This is the core of an astrometric measurement. This relationship between fringe phase (or differential fringe location) and astrometric position goes back to the geometric model of interferometry as shown in Fig.~\ref{fig:21} \citep[see also Sect.~2 in][]{2008NewAR..52..199D}.

The absolute separation between two stars is encoded in the differential fringe location. And the third phase of ASTRA provides the KI with a laser metrology system precise enough to measure $\Delta OPD$ at the 20~nm level, which transforms into a precision better than 100~$\mu$as for stars separated closely enough that the differential atmospheric phase distortions do (nearly) cancel out, i.e. for two stars from within the isopistonic angle.

\section{The ASTRA technology}
\label{sec:4}
In this section we want to present in some more detail the individual systems needed to implement the three phases of ASTRA as described above. This article does not aim at describing the complete design and functionality of each system at an engineering-level. 
But we want to demonstrate the conceptual design of each subsystem to develop a general understanding of the functioning of a modern large aperture dual-field interferometer. 
Such an understanding is required to appreciate the control and calibration work mostly done by {black boxes like} software pipelines and control systems, invisible to the general user. 
But understanding the details of a running system qualitatively also enables the reader to estimate the current technical limits and to foresee which of these limitations might be pushed further in the future by respective developments.

The following sections are arranged with respect to their relation to the three ASTRA phases, which were introduced in Sect.~\ref{sec:3}. Angle tracking (\ref{sec:44}), vibration control of the optical path (\ref{sec:45}) and LGS operation of the interferometer (\ref{sec:47}) will serve all phases, while the field separation (\ref{sec:42}) is only required for the off-axis phase referencing (\ref{sec:32}) and the astrometry (\ref{sec:33}). The interferometric narrow-angle astrometry further requires special systems to monitor the internal differential optical path in dual-field operation (\ref{sec:43}) and the astrometric baseline (\ref{sec:46}).

\subsection{\label{sec:44}Sensitive angle tracking}

\begin{figure}[htp]
\centering
\includegraphics[scale=0.6]{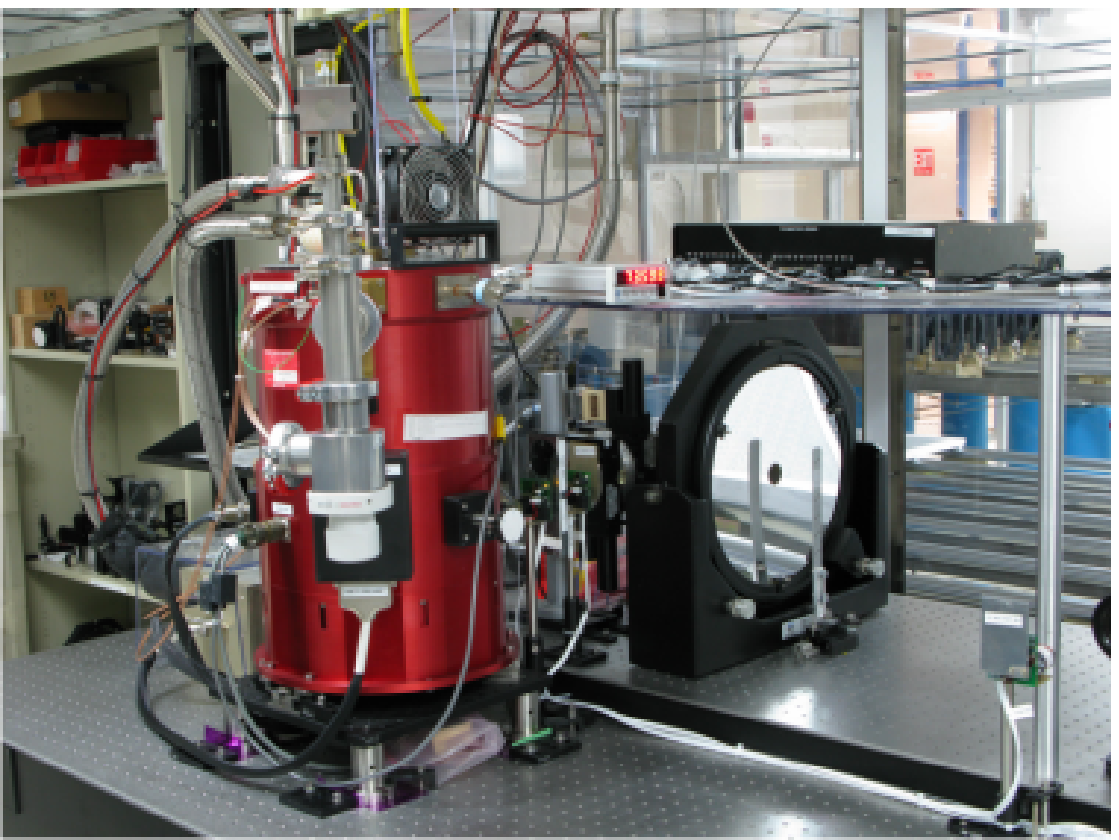}
\includegraphics[scale=0.6]{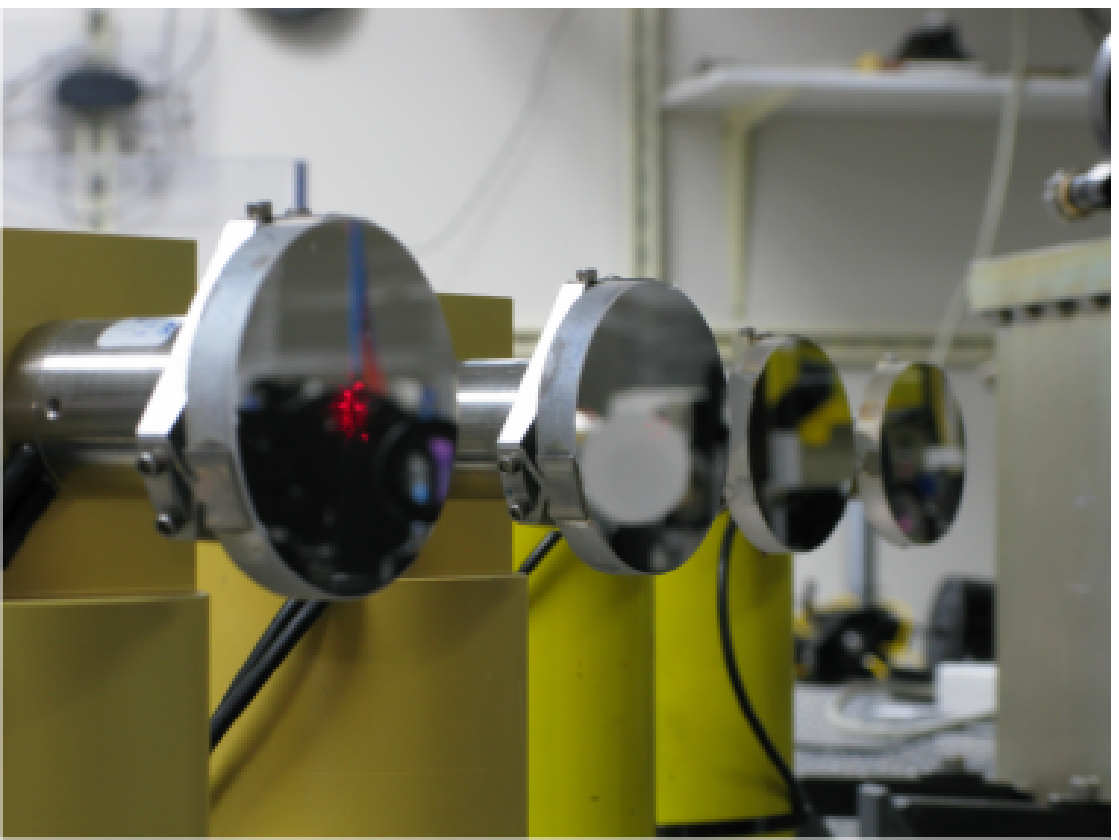}
\caption{\label{fig:6} The Keck Angle-Tracking camera (KAT) in the dewar (left). It commands the fast tip-tilt mirrors which can tilt the collimated beams at a bandwidth of about 100~Hz. This angle-tracking system is essential to continuously achieve high injection ratios in the fringe tracker single mode fiber guaranteeing high fringe SNR.}
\end{figure}

The flux-efficient operation of an interferometric beam-combiner requires a tip-tilt stabilized wavefront. The operation of the KI has shown that turbulence in the long air-filled beam-trains introduce such a tip-tilt variation which is not corrected for by the AO-system since it is mainly introduced by air motion along the optical path between the Nasmyth deck and the basement, i.e. after the AO-system. 
An infrared camera called KAT is in place to monitor this residual tip-tilt and feed back to actuated mirrors to correct for it and stabilize the beam combiner input  (Fig.~\ref{fig:6}). This control loop is currently run typically at 80~Hz. 

A large part of this tip-tilt originates from vibrations of the optics of the interferometer. Currently KAT is monitoring this internal tip-tilt by using the $J$- or $H$-band light of the star which requires stellar magnitudes smaller than $\sim$\,10. Since the dual-field operation foresees observing fainter stars in the phase-referencing mode, an internal light source will be implemented at the telescope to simulate the star light on KAT. This tip-tilt monitor will run at high bandwidths while in addition in a hybrid operation mode the classical on-star tip-tilt correction can work at reduced bandwidth ($\sim\,$1~Hz) to increase the detector sensitivity by the required amount to measure fainter stars, and track the residual drifts.

\subsection{\label{sec:45}Internal OPD stabilization}

Longitudinal vibrations will add artificial piston to the star light and reduce the fringe contrast.
The required vibration control of ASTRA is closely following the concept of the classic KI operation. It is based on a modular design. The primary telescope mirrors are monitored by accelerometers. The additional piston introduced by vibrations along the beam-train are telemetered by a HeNe-laser metrology system. In addition the mount of every single optical element along the optical path is optimized to minimize vibrations. This primary OPD stabilization does not fulfill the task of the {\it differential} dual-field metrology, which is introduced in Sect.~\ref{sec:43} and necessary for the high precision astrometry of ASTRA.

\subsection{\label{sec:47}Two laser guide stars for the interferometer}

\begin{figure}[htp]
\centering
\includegraphics[width=\textwidth]{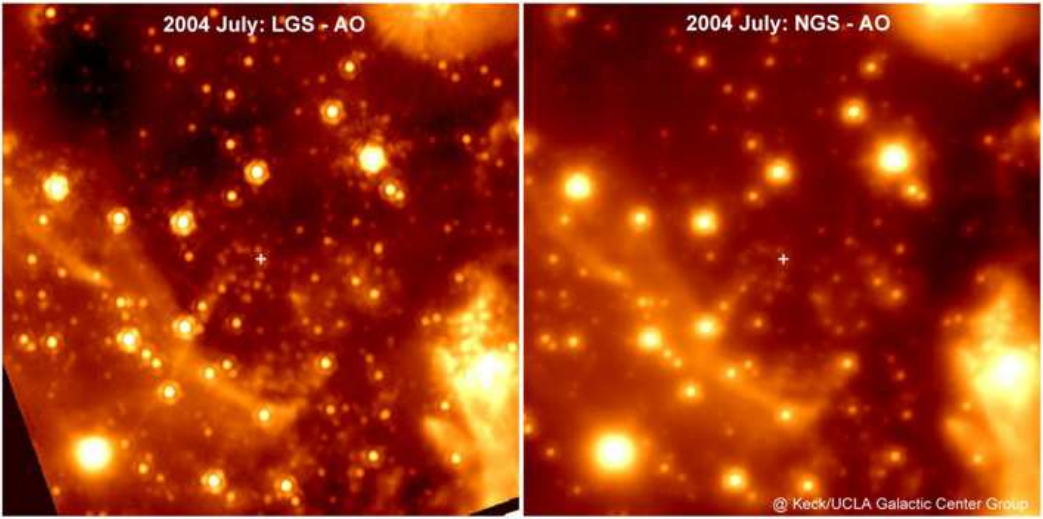}
\caption{\label{fig:15}The UCLA Galactic center group demonstrated the increased efficiency of LGS-AO assisted Keck observations on stars in the Galactic center which appear very red due to the high interstellar extinction. The closest natural guide star (NGS) is a foreground star about 20~arcsec away from the very center. The location of the central massive black hole is marked with a white cross in these $L^\prime$-band images, which are 7.5$^{\prime \prime}$ across. The Strehl ratio is increased by a factor of 2 by the LGS \citep[see][for further details on the NGS vs. LGS comparison of Galactic center observations]{2005ApJ...635.1087G}.}
\end{figure}

AO-correction of the incoming wavefronts is crucial for the KI to enable high fringe SNR in particular when the correlated flux is low due to a source extension resolved by the interferometer. 
An adaptive optics system is needed to ensure that as much light as possible is coupled into the fiber by flattening the wavefront and correcting for atmospheric tip-tilt (see also Sect.~\ref{sec:44}).
The KI is only operated with closed AO-loops at both telescopes. 
The Keck AO-system (as most others) is equipped with a visible wave-front sensor whereas the interferometric measurement is done in the near-infrared. Most stellar sources are bright enough to fulfill both sensitivity constraints: in the visible for the AO wavefront sensor, and in the near-infrared (NIR) for the fringe measurement. 
But the AO-limit of $R\,\sim\,12$ prevents on-axis AO-corrected KI-observation of red objects which would be bright enough in the NIR: dust enshrouded YSO's and evolved stars, sources in the Galactic center due to the high amount of interstellar extinction, and some active galactic nuclei (AGN).

To enable KI-observation of such red targets, and to make the broadest use of the ASTRA upgrade, a second laser guide star system (LGS) system is currently implemented to work in parallel to the original LGS at the other telescope. Once available, interferometric observations with both telescopes' AO-loops locked on an artificial laser beacon are possible. Such LGS-IF operation of the KI is expected to  stabilize the performance of the interferometric measurement of intrinsically faint objects similar to the gains achieved with single telescope LGS-assisted observations \citep[and Fig.~\ref{fig:15}]{2004SPIE.5490..321B}.
First on-sky tests of LGS-IF operation are foreseen for early 2010.

\subsection{\label{sec:42}Field separation}

\begin{figure}[htp]
\centering
\includegraphics[width=\textwidth]{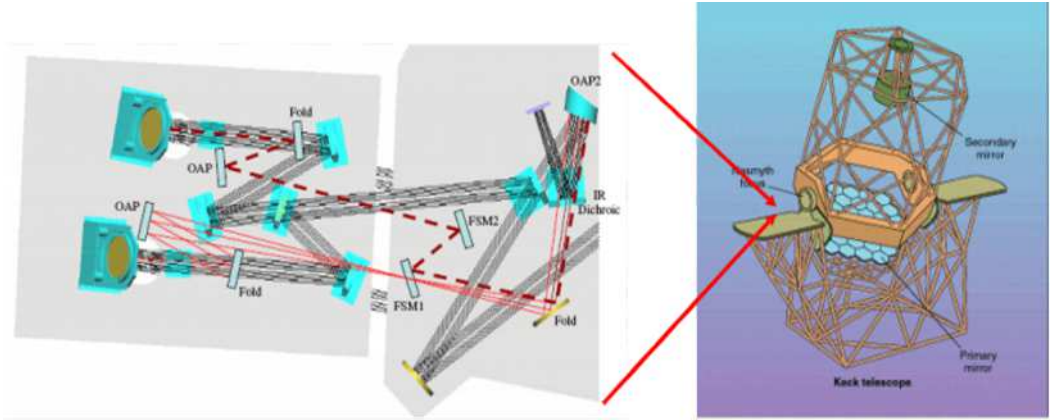}
\caption{\label{fig:3}Conceptual design of the ASTRA dual-field facility. It is locate in the Nasmyth focus of each telescope (right panel). 
The red lines (solid and dashed) are representing the optical path of the primary and secondary star, respectively. A central uncoated area in the optic called FSM1 transmit the primary star light, and reflects the surrounding light. FSM1\&2 are movable to select the location of the secondary star to be relayed to the beam train after collimation in an off-axis parabola (OAP).}
\end{figure}

The field separation in the Nasmyth foci of the telescopes is a crucial process at the heart of the ASTRA project (Fig.~\ref{fig:3}). 
The ASTRA design concept includes the use of an annular mirror which reflects an {\it off-axis} field of 60~arcsec diameter, centered on the {\it on-axis} phase-reference star. 
The size of the usable off-axis field is given by the isopistonic angle. 
After field separation the re-collimated light enters the underground beam trains and delay lines. 
The annular mirror will be actuated to select the right off-axis star for the second beam train. For this blind step to work, good differential coordinates of sub-arcsec precision are required. These are typically provided by an AO-assisted $K$-band image of a large telescope. 

\subsection{\label{sec:43}Internal differential dual-field metrology}

Following Eq.~\ref{eq:1}, the astrometry requires to measure precisely the {\it differential} optical path difference $\Delta OPD$, created by the differential phase of the two stars (Fig.~\ref{fig:21}).
One of the technical challenges is that the light of the two objects will travel along different optical paths after the field separation. A static difference in the optical paths could be measured once and calibrated out. But to guarantee that this difference is continuously known down to an accuracy of a few nm, an internal laser metrology system will be implemented.
The task is to track the differential internal optical path differences which will vary with time due to changing conditions in the transmitted air and mirror position.
To minimize dispersion induced misinterpretation of the metrological signal the wavelength of the metrology laser will be in the infrared, close to the science wavelengths. 

\subsection{\label{sec:46}Monitoring the astrometric baseline}

\begin{figure}[htp]
\centering
\includegraphics[scale=0.8]{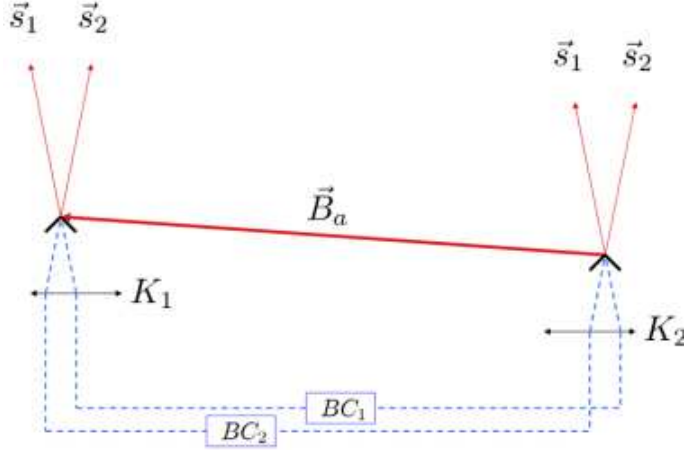}
\caption{\label{fig:5} A sketch showing a typical astrometric baseline. K1 and K2 denote the primary apertures of the Keck telescopes, they shall include the pivot point of the telescopes. The black corners represent the reflection corner cubes of the internal differential metrology system, projected in the primary space of the telescopes. The astrometric baseline $\vec{B}_{\rm a}$ can always be defined as the connection between these conjugated locations of the corner cubes if the corner cubes cannot be positioned exactly at the pivot points. 
}
\end{figure}

Having a goal of sub-100~$\mu$as precision for the astrometry not only requires a very precise knowledge of the differential $OPD$.
Also the baseline vector $\vec{B}$ is required to be known at the 100~$\mu$m precision level. 
A careful investigation of the instrumentation concept, namely the internal differential dual-field metrology, reveals a particularity of ASTRA-like interferometric astrometry. 
The {\it astrometric} baseline, needed to derive the angle between the two stars from the differential phase (or $OPD$) measurement (Eq.~\ref{eq:1}), is not simply the separation vector between the two telescopes. 
Since the differential phase measurement relies on the internal differential dual-field metrology, 
the astrometric baseline is given by the difference vector of the end-points of this metrology, calculated in the primary space of the telescope (Fig.~\ref{fig:5}, \footnote{The metrology endpoints are physically located in the beam-compressed space after the primary mirrors. This requires to conjugate the endpoints back into the 10~m primary space to understand or predict the $\Delta OPD$ in relation to the star separation angle.}). The astrometric baseline is equivalent to the plate scale in imaging astrometry.

For opto-mechanical reasons these endpoints are not exactly coincident with the pivot points of the telescopes, which are usually used to define the interferometric (or so-called wide-angle) baseline vector. The ASTRA internal metrology will end at the telescope focus and not at the pivot points which are close to the tertiary mirrors. 
In the three-dimensional telescope primary space the astrometric baseline is only identical with the wide-angle baseline if the non-monitored paths between pivot point and the endpoints of the metrology system are identical for both telescopes. Otherwise the non-monitored path difference has to be accounted for (see Fig.~\ref{fig:5}).

Experimental data show that the 10~m Keck primaries and their mounts are moving on the order of the required precision when slewed significantly. As part of the ASTRA upgrade we plan to monitor this mostly random telescope runout with an imaging system to understand the effective change of the baseline when slewing between calibrator and science targets. It is further planned to investigate if it will be necessary to monitor the actual distance between the pivot point and the endpoints of the metrology. A change of this distance can originate from telescope flexure over the night.

\section{KI-ASTRA astrometry science}
\label{sec:5}

\begin{figure}[htp]
\centering
\includegraphics[width=0.6\textwidth]{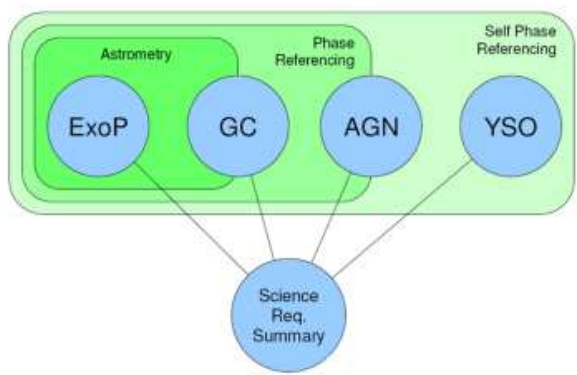}
\caption{\label{fig:8}Schematic overview of how the science cases drive the instrument requirements of ASTRA. The disks around young stellar objects (YSO) are typically bright enough to be studied in self-phase-referencing, while the dual-field phase-referencing is required to go deeper and to study the circumnuclear emission of AGN and stars in the Galactic center (GC). Finally the astrometry is designed for both exo-planet and GC observations as outlined in the Sec.~\ref{sec:5}. The astrometry relies on completion of the previous SPR- and DFPR-phases of the ASTRA upgrade.}
\end{figure}

KI-ASTRA upgrade offers a variety of new observing capabilities, including higher spectral resolution, better sensitivity and a precision of differential astrometry beyond current single telescope limits (Sect.~\ref{sec:3}). 
This enables a large number of new science cases ranging from studying the stellar reflex motion of exo-planet systems, and the astrophysical properties of planet-forming disks over classical interferometry science at unprecedented sensitivity to a systematic study of AGN with the phase-referencing technology (see Fig.~\ref{fig:8}). 

In this series we focus on the science cases which drive the development of the astrometric mode as outlined in the following sections. 
Be reminded of the general requirements and assets for such an astrometry science case
\begin{itemize}
\item  Two stars / targets per observation only; while the astrometric precision will surpass single telescope imaging limits, the efficiency of {\it "measured separations per observing time"} is low and not applicable for complex cluster astrometry with hundreds of stars to be monitored
\item High precision is achieved with {\it only} two stars whereas AO-assisted single telescope imaging requires a large number of stars to reach an astrometric accuracy of $\sim \,$100~$\mu$as, to calibrate for the distortion over the field-of-view \citep{2008preprintC}
\item Target separation is bound to be larger than about 1~arcsec for technical reasons, and smaller than 30~arcsec due to the atmospheric limitations
\item At least one target needs to be brighter then 10~mag in $K$ to track fringes on, the fainter object shall be brighter than 15~mag; the exact gain due to phase referencing depends on the final stability on the instrument, and might be higher if minute long coherent integrations can be achieved.
\item KI elevation limit is $\sim\,$38~degr, limiting the observability of Southern targets
\end{itemize}
\subsection{Precision astrometry to weigh stellar companions} 
The precise astrometry will help to increase the precision of binary kinematic studies, which are an efficient way to measure the involved masses if the distance to the binary system is known. 
In particular interferometric binary measurements have been shown to deliver crucial mass estimates for pre-main sequence stars where the stellar evolution is still poorly understood \citep{2007ApJ...670.1214B}. Observational mass estimates help to gauge the models of star formation. The increased sensitivity will increase the number of accessible targets and stellar types. E.g. brown dwarfs with typical brightnesses of $K>\,$10 are now for the first time within interferometric reach. ASTRA observations will help to better understand the formation and evolution of these  transition objects between planets and ordinary main-sequence dwarfs. 

\subsection{Studying exo-planetary kinematics}
Radial velocity (RV) measurements are so far the most efficient way to {\it detect} planetary systems \citep{2008NewAR..52..154S}. But RV-experiments miss the orbital inclination unless rare edge-on transiting-planet systems are observed. 
RV-planet surveys reveal many systems in which the stellar reflex motion can be detected with differential astrometry at the 100~$\mu$m accuracy level, that is at the level of ASTRA precision.
Interferometric astrometry offers a means of measuring orbital inclinations with an accuracy of a few percent. 
A focus of the ASTRA planet science case are multiple-planet systems.
Co-planarity and the understanding of planet migration, scattering, and capturing processes due to planet-planet interaction have long awaited an experimental input to challenge theoretical ideas and models \citep{2007lyot.confQ..21M}.
To enlarge the parameter space of planet detection toward less massive planets, beyond the reach of current RV-surveys, the 50~$\mu$as accuracy level has to be reached, one of the final goals of the ASTRA-project \citep[see Fig.~3 in][]{2002ApJ...574..426E}.

\subsection{Astrometry in the Galactic center}

Another area of research, where ASTRA-astrometry will have a significant impact, is the central parsec of our Galaxy (GC). 
Near diffraction limited imaging, since most recently assisted by laser guide stars,  have led to the first doubtless detection of a massive black hole at the gravitational center of the Milky Way via astrometric monitoring of the Keplerian motion of stars orbiting the black hole. Higher astrometric precision will help to further narrow down the uncertainties of the black hole mass ($4.5\,\pm\,0.4\times 10^6\,M_\odot$) and the solar distance to the GC \citep[$R_\odot\,=\,8.4\,\pm\,0.4\,{\rm kpc}$, see][and references therein]{2008ApJ...G}.
Currently the best constraints derive from the closest and brightest of the orbiting stars, S0-2 ($K\,\sim\,14$). 
It can be observed against the nearby evolved maser stars, only 5-10~arcsec away from the GC, bright enough to deliver an astrometric phase reference of an ASTRA astrometry observation \citep[see Fig.~15 in ][]{2008ApJ...G}. 

The differential absolute astrometric measurement of the orbiting star will give for the first time an independent estimation of the accuracy of the imaging astrometry in that field. 
The higher angular resolution of the interferometer will help estimate astrometric biases of the single telescope measurements, such as source confusion or stellar multiplicity. 
Currently these biases prevent from increasing the measurement accuracy of the properties of the black hole by using more orbiting stars in addition to S0-2.
The final limiting magnitude of the ASTRA-astrometry mode will be crucial to observe numerous stars in the GC due to the high amount of interstellar extinction along the line-of-sight to the GC through the disk of the Galaxy. Further chances and challenges of the ASTRA-GC science case have recently been discussed elsewhere \citep{2008SPIE.7013P}.

\section{Summary}
\label{sec:6}

We presented the concepts of the ASTRA sensitivity upgrade of the Keck Interferometer. A first successful commissioning run of the SPR-mode has been conducted in spring 2008. The various subsystems needed to implement the next phases, dual-field phase-referencing and the interferometric astrometry, have been introduced on a conceptual level. The gradual integration of ASTRA and its subsystems over the next two years into the running infrastructure of the KI will enable to offer the new instrumentation to the observing community as soon as implemented. 
We named exo-planet research, precision astrometry of binary systems, and the improvement of measurements of the central dark mass concentration at the heart of the Milky Way as the primary science cases for the astrometric mode of ASTRA. 

\section*{Acknowledgments}
The W.\,M.~Keck Observatory is operated as a scientific partnership among the California Institute of Technology, the University of California, and the National Aeronautics and Space Administration. The Observatory was made possible by the generous financial support of the W. M. Keck Foundation.
The authors wish to recognize and acknowledge the very significant cultural role and reverence that the summit of Mauna Kea has always had within the indigenous Hawaiian community.  We are most fortunate to have the opportunity to conduct observations from this mountain. The Keck Interferometer is funded by the National Aeronautics and Space Administration. The ASTRA upgrade is funded by the Major Research Instrumentation program of the National Science Foundation.




\end{document}